\documentclass[aps,a4paper,showpacs,twocolumn]{revtex4}

\usepackage{epsfig}
\usepackage{amsmath,amssymb,color}
\usepackage[english]{babel}

\parskip=\medskipamount

\newcommand{\eq}[1]{(\ref{#1})}
\newcommand{\fig}[1]{Fig.\ref{#1}}

\newcommand{\be}{\begin{equation}}
\newcommand{\ee}{\end{equation}}

\newcommand{\barr}{\begin{array}}
\newcommand{\earr}{\end{array}}

\newcommand{\beqn}{\begin{eqnarray}}
\newcommand{\eeqn}{\end{eqnarray}}

\newcommand{\bs}{\begin{subequations}}
\newcommand{\es}{\end{subequations}}

\newcommand{\bw}{\begin{widetext}}
\newcommand{\ew}{\end{widetext}}

\newcommand\disp{\displaystyle}

\begin{document}

\title{Anomalous diffusion in fractal globules}

\author{M.V. Tamm$^{1,2}$, L.I. Nazarov $^1$, A.A. Gavrilov $^{1,3}$, A.V. Chertovich $^1$}

\affiliation{$^1$Physics Department, Moscow State University, 119991, Moscow, Russia \\
$^2$ Department of Applied Mathematics, National Research University Higher School of Economics, 101000, Moscow, Russia\\
$^3$ Institute for Advanced Energy Related Nanomaterials, University of Ulm, D-89069, Ulm, Germany}

\date{\today}

\begin{abstract}

The fractal globule state is a popular model for describing
chromatin packing in eukaryotic nuclei. Here we provide a scaling
theory and dissipative particle dynamics (DPD) computer simulation for
the thermal motion of monomers in the fractal globule state. Simulations starting from different entanglement-free
initial states show good convergence which provides evidence supporting the existence of unique metastable fractal globule state. We show monomer motion in this state to be sub-diffusive described by $\langle X^2 (t)\rangle \sim t^{\alpha_F}$ with $\alpha_F$ close to 0.4. This result is in good agreement with existing experimental data on the chromatin dynamics which makes an additional argument in support of the fractal globule model of chromatin packing.    

\end{abstract}


\maketitle

The question of how genetic material is packed inside an eukaryotic nucleus is one of the most challenging
in contemporary molecular biology. This packing, apart from being very compact, has some striking
biological properties including existence of distinct chromosome territories, easy unentanglement of
chromosomes and chromosome parts (needed in preparation to mitosis, and during transcription), and ability of
different parts of the genome to find each other in space strikingly fast in e.g. so-called promoter-enhancer
interactions. All these properties are very untypical for the compact states of generic synthetic polymers
(known as equilibrium globular state in classical polymer physics, see, e.g., \cite{deGennes,GrKhokh}).
Indeed, e.g., human \emph{chromatin} fiber (that is to say, a composite polymer fiber consisting of dsDNA and
associated \emph{histone} proteins\cite{balberts}) is so long that an equilibrium polymer globule made of it
would be too entangled to perform any biological functions on any reasonable
time-scale \cite{rosa_everaers}.

The theories proposed to explain the chromatin packing tend to either base on some \emph{ad hoc} biological
mechanism of stabilization \cite{loop1,loop2,loop3,loop4,loop5,pnas} or argue that topological interactions
prevent the chromatin chain from entangling itself on biological timescales
\cite{grosb_conj,rosa_everaers,Lieberman,MirnyChrom}. The latter point of view relies on the analogy
between chromatin state and other topologically-governed polymer states, such as fractal (crumpled) globules
\cite{GNS,grosb_conj,nechaev1,nechaev2}, and melts of non-concatenated polymer rings
\cite{cates,sakaue1,sakaue2,HalversonPRL,GrosbSoft}. In recent years, the data obtained via novel
experimental techniques to study genome structure, in particular FISH \cite{FISH} and Hi-C
\cite{dekker,Lieberman,naumova} methods seem to provide data supporting the topological fractal globule
approach.

For the detailed overview of the state of the field we address the reader to a recent
review\cite{grosbreview}, here we provide a brief summary of those presumed static properties of
chromatin packing which we use in what follows. First, fractal globule model assumes that chromatin fiber forms a
compact fractal state with dimensionality 3, i.e., on all length scales the typical spatial
distance $R$ between monomers depends on genomic distance $n$ between them as $R\sim n^{1/d_f} =
n^{1/3}$. Second, there are no entanglements in the fractal globule contrary to the equilibrium one. Because
of that, parts of chromatin can easily fold out from the fractal globule conformation and form extended loops,
and then retract back to refold into the dense state. Third, fractal globule has a
distinct territorial organization: parts of the genome close to each other along the chromosome are close to
each other in space as well. These properties are dictated by the absence of knots on the chromatin chains and topological entanglements between them and are, therefore, shared between linear polymers in unentangled state and non-concatenated unknotted rings; the difference is that while for rings the described fractal state is equilibrium, for linear chains it is but metastable, although it is supposed to be relatively long-lived. The question
of how to prepare a fractal state with long-living stable properties is still a matter of debate, many
different algorithms to prepare a fractal globule in computer simulations have been suggested
\cite{Lieberman,Fractalglobule_modeling,Rosa_Everaers_new,Smrek,Chertov_Kos,NNT_prep}, most of them appear to be evolving in time rather rapidly when the simulation starts. Here we use two different algorithms to
prepare initial fractal states, then anneal them for some time before starting measurements. The
observed convergence of the results obtained from two different initial states suggests that there indeed
exists a unique metastable fractal globule state corresponding to a partial equilibrium of the polymer chain
given the absence of topological entanglements.

Dynamics of a fractal globule state, which is a focus of this paper, has been less studied so far. Clearly, self-diffusion in the fractal globule should be faster
than in the equilibrium one due to the absence of entanglements\cite{kremer}. Sometimes
\cite{Fractalglobule_modeling,Imak_Mirny} the Rouse dynamics of the fractal state is assumed in order to
estimate the relaxation times of the chain as a whole, while explicit measurements (e.g., computer
simulations of non-concatenated rings\cite{HalvJCP}, experiments on the dynamics of unknotted ring bacterium genomes \cite{mcl} and on the telomeres in the nuclei
\cite{kepten1,kepten2,kepten3}) suggest a slower than Rouse dynamics. Indeed, the discrepancy from Rouse
theory is to be expected since it relies heavily on the absence of interactions between monomers which are not immediate neighbors along the chain \cite{deGennes}, and cannot be directly applied to the fractal globule which is actually stabilized by this interactions. Recently, theoretical approach to generalize the Rouse model to produce different scaling exponents was suggested \cite{amitai} but without any discussion of what particular exponent one should choose in physically relevant situation
\footnote{Note also papers \cite{LongDiff1,LongDiff2} where different but related question of the probe particle diffusion in chromatin matrix is investigated. When this paper was already submitted for publication we became aware of paper  \cite{grosb_diff} were fractal globule dynamics is addressed. Results of \cite{grosb_diff} are substantially different from ours (they predict scaling exponent $\alpha_F \simeq 0.26$), we believe that their result is due to a mistake in identifying all possible elementary movements of a chain}. In
what follows we present a scaling theory and computer simulations of the self-diffusion in a fractal globule
state resulting in a subdiffusive motion with an exponent similar to one observed experimentally in
\cite{mcl, kepten1,kepten2,kepten3}).

We start with Rouse model, which is the simplest model of the dynamics of an unentangled polymer. In the continuous limit the conformation of the Rouse chain $X(s,t)$ (here $X$ is the spatial coordinate, $s$ is a coordinate along the chain, and $t$ is time) satisfies equation\cite{deGennes, GrKhokh}
\be
\frac{\partial X(s,t)}{\partial t} = \lambda \frac{\partial^2 X(s,t)}{\partial s^2} + \xi (s,t)
\label{Rouse}
\ee
where $\lambda$ is some coefficient, $\xi$ is white thermal noise delta-correlated in space and time. This equation has a stationary solution, which is a Gaussian measure over all trajectories $X(s,t)$. In what follows we restrict ourselves to discussing very long chains (or, equivalently, relatively short times) which makes the boundary conditions coupled to Eq. \eq{Rouse} irrelevant for internal monomers.

Eq. \eq{Rouse} neglects any interactions between monomers not immediately adjacent along the chain, and therefore it cannot be directly applied to non-Gaussian equilibrium or metastable states of a polymer chain, which are stabilized by volume interactions. Many different generalizations of Eq. \eq{Rouse} are possible, e.g. by introducing fractional derivatives \cite{amitai, metzler_review} or by introducing correlations into the noise term \cite{stanford}. It is not clear which particular generalization is most valid microscopically for the  fractal globule, so instead of modifying Eq. \eq{Rouse} we rely below on a more general scaling argument. Proceeding this way we loose the detailed information about the statistics of the monomer self-diffusion, but are able at least to recover the scaling exponent of the self-diffusion.

For Rouse model the scaling argument goes as follows. Let $x(s,t)$ be a stationary solution of Eq. \eq{Rouse}. Then for any given time $t$ and two positions along the chain $s_1,s_2$
\be
<(x(s_1,t)-x(s_2,t))^2> \sim |s_1-s_2|,
\label{Rouse_dist}
\ee
where triangular brackets correspond to averaging over stationary solutions of Eq. \eq{Rouse} Assume now that as time goes on the monomer displacement grows as
\be
<(x(s,t+\tau)-x(s,t))^2> \sim (\tau)^{\alpha}
\label{Rouse_displ} 
\ee
with some unknown $\alpha$. Since the chain is connected and Eq. \eq{Rouse_dist} holds at any given time, parts of the chain of length $\delta s (\tau) = |s_1-s_2| \sim (\tau)^{\alpha}$ are obliged to move collectively at a timescale $\tau$. Moreover, if all monomers in this chain fragment experience independent random forces from the solvent, the collective effective diffusion constant of such a fragment is 
\be 
D(\delta s) \sim D_0 / \delta s = D_0 \tau^{-\alpha},
\label{Rouse_diff}
\ee
where $D_0$ is a microscopic diffusion constant\footnote{Let us emphasize that $\tau$ dependence of the diffusion constant in \eq{Rouse_diff} is not to be confused with the time dependence of the diffusion constant in the so-called scaled Brownian motion, which in our notation would correspond to dependence on $t$ variable. Indeed, Rouse model for an infinite chain  corresponds not to scaled Brownian motion but to fractional Brownian motion, which has very different statistical and ergodic properties. For a comprehensive review of different models of anomalous diffusion see \cite{metzler_review}}. Combining Eqs. \eq{Rouse_displ} and \eq{Rouse_diff} one recovers the well-known result 
\be
\begin{array}{l}
<(x(s,t+\tau)-x(s,t))^2> \sim (\tau)^{\alpha} \sim D(\delta s) \tau \sim \tau^{1-\alpha}; \medskip \\ \alpha_{R} = 1/2, 
\end{array}
\label{Rouse_alpha} 
\ee
where we introduced notation $\alpha_R$ for the scaling exponent of the Rouse model.

This scaling reasoning is much easier to generalize for the fractal globule case than Eq. \eq{Rouse} itself. Indeed, the princaipal change is the statistic of the state we consider (recall that we are only considering time scales much shorter than chain entanglement time, so we assume that fractal globule state can be treated as stationary). This corresponds to replacing Eq. \eq{Rouse_dist} with 
\be
<(x(s_1,t)-x(s_2,t))^2> \sim |s_1-s_2|^{2/d_f},
\label{fractal_dist}
\ee
where $d_f$ is a fractal dimensionality of the state under consideration, $d_f=3$ for a fractal globule. The chain connectivity argument still holds, and the size of a collectively moving domain scales now as $\delta s (\tau) \sim (\tau)^{\alpha d_f/2}$. If  the random forces acting on monomers are still independent, the resulting scaling exponent of a fractal globule $\alpha_F$ is
\be
\begin{array}{l}
<(x(s,t+\tau)-x(s,t))^2> \sim (\tau)^{\alpha} \sim \frac{D_0}{\delta s} \tau \sim \tau^{1-\alpha d_f/2}; \medskip \\ \alpha_{F} = \frac{2}{2+d_f}=2/5, 
\end{array}
\label{Rouse_alpha} 
\ee
Similar predictions for the self-diffusion in swollen polymer coils has been coined previously, see, e.g., \cite{kardar}. In \cite{suppmat} we argue that allowing for hydrodynamic interactions should make the forces acting on different monomers correlated which will speed up the diffusion. We show, however, that this effect is expected to be small, only shifting $\alpha_{F}$ to around 0.42. 

\begin{figure*}
\epsfig{file=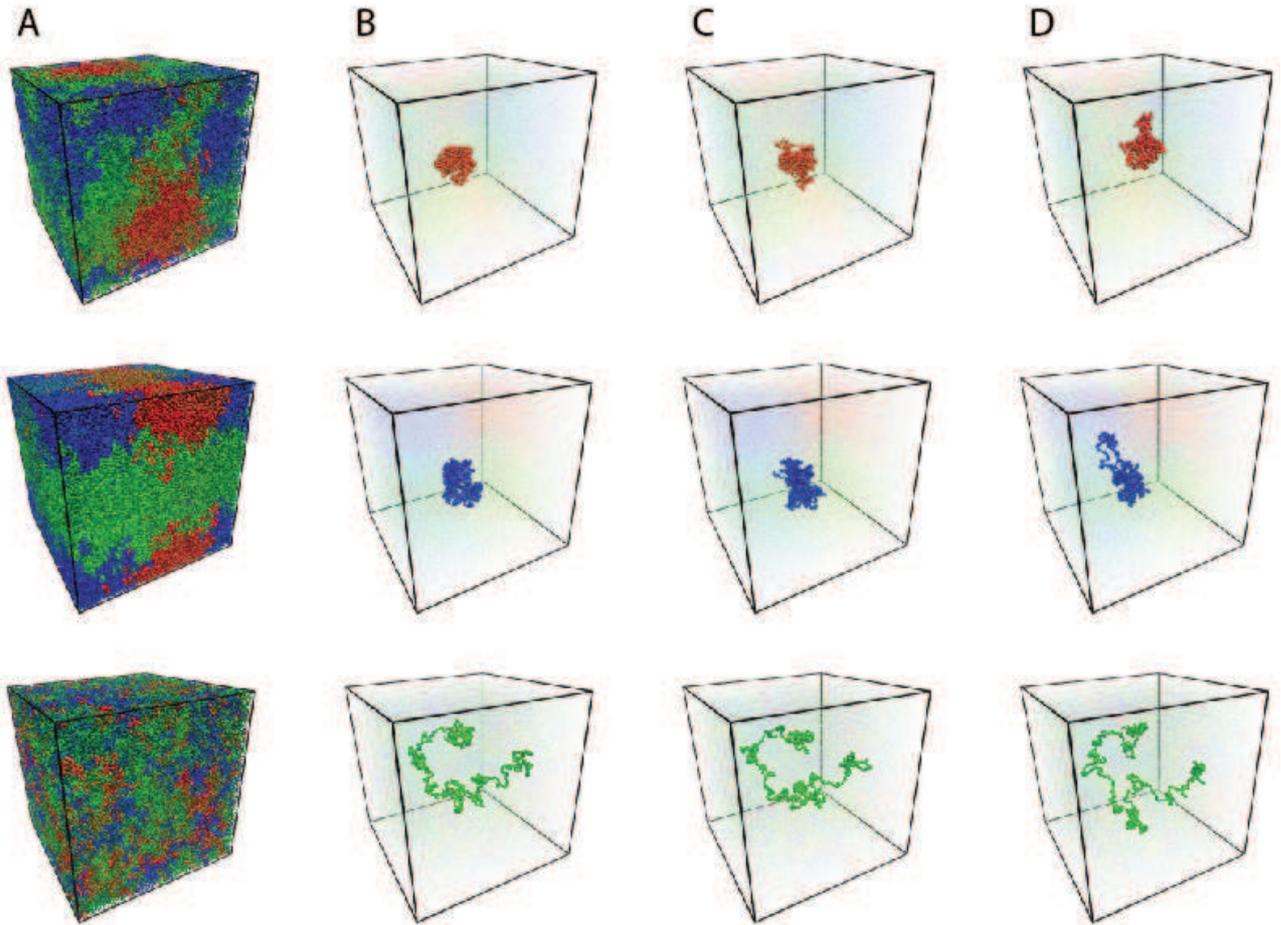, width=17 cm} \caption{ The snapshots of globule conformations: random fractal (top), Moore (middle), and Gaussian (bottom) globules. A): General view of the modeling cell after initial annealing. Chains are gradiently colored from blue to red. B)-D): The evolution of a 1000-monomer subchain conformation: (B) initial conformation at the start of measurement, (C) after $2^{18} \approx 2.5 \times 10^5$ DPD steps, (D) after $2^{26}
\approx 6.5 \times 10^7$ DPD steps. The cube on the figure corresponds to the whole simulation box and has the size 46x46x46 DPD length units.}
\label{fig_snapshot}
\end{figure*}

The natural state of comparison for a fractal globule is a usual entangled equilibrium globule, where self-diffusion of monomers is described by the Rouse exponent $\alpha_R = 1/2$ only on short time scales, when displacement is smaller than the typical size of the entanglement blob. For larger time and length scales the entanglements pay crucial role and the scaling theory \cite{GrKhokh} predicts $\alpha_{\mbox{ent}} = 1/4$.

\begin{figure}
\epsfig{file=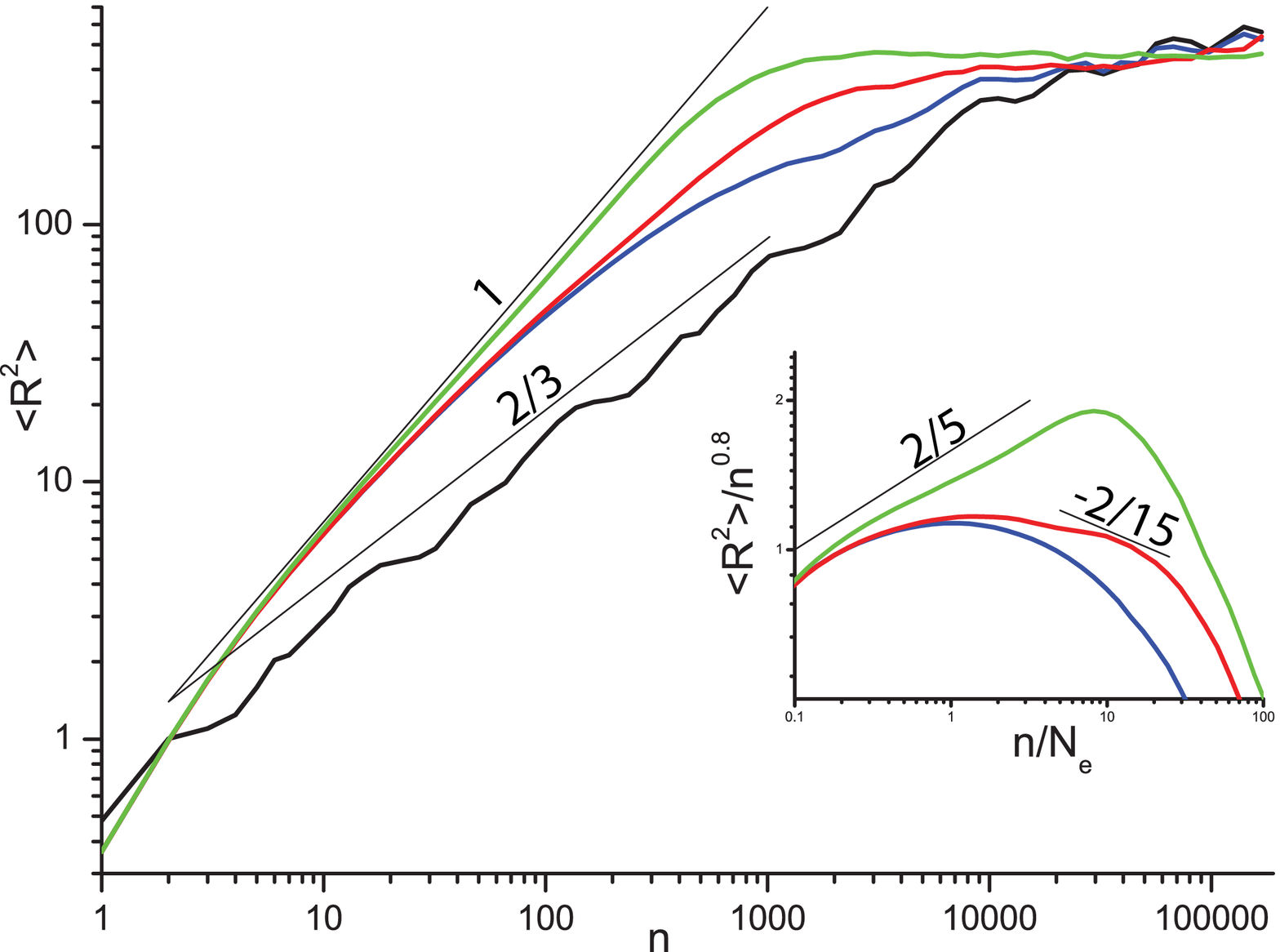, width=8.5 cm} \caption{ Mean-square distance $\langle R^2 \rangle$ between monomers as a function of genomic distance $n$. Gaussian (green) and random fractal (red) states are stable on the modeling timescale (see Figure 2 in
\cite{suppmat}). Initial Moore state (black) relaxes after annealing to the blue curve, approaching the random
fractal state. Inset shows same plots in $\langle R^2 \rangle n^{-0.8}$, $n/N_e$ coordinates used
in \cite{HalversonPRL}.}
\label{fig_characterization}
\end{figure}

To check the predictions of the scaling theory we held out extensive computer simulations using the
dissipative particle dynamics (DPD) technique which is known \cite{groot,nikunen} to correctly reflect
dynamics of dense polymer systems. The polymer model we use consists of
renormalized monomers with the size of order of the chromatin persistence length, corresponding DPD time step is of order $1 n sec$ or more (see \cite{suppmat} for more details). Volume interactions between the monomers are chosen to guarantee the absence of chain self-intersections, the entanglement length is $N_e\approx 50 \pm 5$ monomer units \footnote{We use the method described in\cite{karatrantos} to determine $N_e$, the referred value corresponds to the so-called s-coil definition of
$N_e$.}. The modelled chains have $N=2^{18}=262144$ units confined in a cubic volume with periodic boundary conditions. In a chain that long ($N/N_e \simeq 5000$)
the equilibration time by far exceeds the times accessible in computer simulation, so the
choice of starting configurations plays a significant role. Here we provide a short outline of how we
construct and prepare the initial states, addressing the reader to \cite{suppmat} for further details.

The first initial state we use is a randomized Moore curve similar to that described
in\cite{Fractalglobule_modeling}, it has a very distinct domain structure with flat domain walls. The second
initial state is generated by a mechanism which we call ``conformation-dependent polymerization in poor
solvent''. This algorithm, which, for the best of our knowledge, has never been suggested before, is
constructing the chain conformation by consecutively adding monomer units in a way that they tend strongly to
stick to the already existing part of the chain. In \cite{suppmat} we show that the resulting conformations
show exactly the statistical characteristics expected from fractal globules, while a full account of this new
algorithm will be given in \cite{NNT_prep}. In what follows, for brevity we call the globule prepared by the
randomized Moore algorithm ``Moore'', and one prepared by the conformation-dependent polymerization ``random
fractal''. As a control sample we use a standard equilibrium globule which we call ``Gaussian''.

Prior to the diffusion measurements all three initial states are annealed for $\tau = 3.2\times10^7$ modeling
steps. The statistical properties of random fractal and Gaussian globule does not change visibly during the
annealing time, while the Moore globule is evolving with domain walls roughening and its statistical
characteristics (e.g., dependence of the spatial distance between monomers on the genomic distance
$\langle r^2(n)\rangle$, see \cite{suppmat}) approaching those for the random fractal globule state.

Snapshots of conformations annealed from different initial states are shown in \fig{fig_snapshot}. In fractal states, contrary to the Gaussian one, fragments close along the chain tend to form domains of the same color. The states are
further characterized in \fig{fig_characterization}. The fractal globule curve appears very similar (but
for the saturation at large $n$ due to the finite size effects) to the universal spatial size-length curve for
unentangled rings discussed in\cite{HalversonPRL,johner}. $R^2 (n)$ for the Moore state seems to approach the
fractal globule curve with growing modeling time suggesting the existence of a unique metastable fractal globule state.
Fractal globules prepared by two different techniques are significantly different at first, but converge with growing simulation time, making the results obtained after annealing unsensitive to the details of the initial state.

\begin{figure}
\epsfig{file=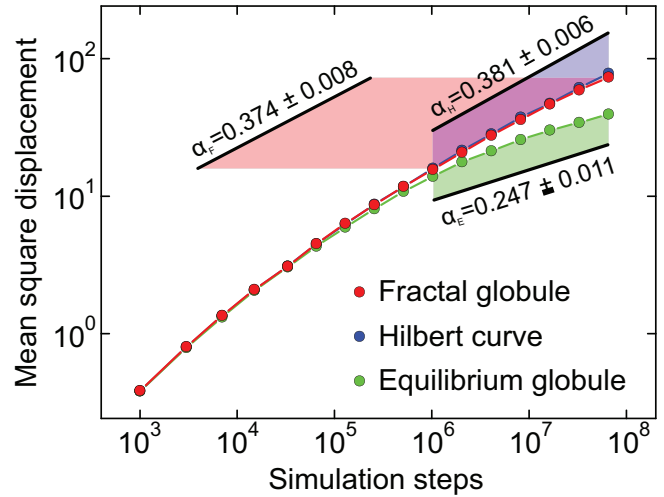, width=8.5 cm} \caption{ Average mean-square
displacements of monomer units as a function of time. Globules starting from an irregular fractal (red),
regular Gilbert (green), and equilibrium globule (blue) initial conformations. All conformations where
annealed for $2^{25}=3.2\times10^7$ timesteps before the measurements started.}
\label{fig_common}
\end{figure}

Monomer spatial displacement was measured for $t = 6.5\times10^7$ DPD time steps after the annealing (corresponding to $\sim 0.1 sec$ on the real time scale), with results shown in figure
\ref{fig_common}. Impressively, mean-square displacement for the random fractal and Moore initial states is indistinguishable within the measurement error. As expected, it is slower than in the Gaussian state: the observed scaling exponent for Gaussian globule $\alpha_G$ is
fairly close to $\alpha_{\mbox{ent}}=1/4$ predicted by the reptation model, while for the fractal globule one
gets $\alpha_F^{exp} \approx 0.38$ which clearly is above $\alpha_{\mbox{ent}}=1/4$ and below the Rouse exponent
$\alpha_R=1/2$, fairly close to our theoretical prediction $\alpha_F^{th}= 0.4..0.42$. 
Both our simulations and results known from the literature for computer simulation \cite{HalvJCP,hur} and experiment \cite{kepten1,kepten2,kepten3,mcl} of similar unknotted polymer systems give scaling exponents similar but slightly below our theoretical estimate. We expect this discrepancy between theory and simulation to be due primary to the fluctuation effects. We also examined the distributions of monomer displacements at various times for all three initial states \cite{suppmat}. For both fractal states the monomer displacement distributions stay Gaussian at all times despite the mean-square displacement growing subdiffusively, a behavior typical for fractional Brownian motion\cite{metzler_review}. In turn, distribution of monomer displacements in the equilibrium globule shows visible deviations from the normal distribution. 

The scaling theory introduced above can be used to estimate the first passage time for two parts of a chromatin chain (e.g., the loci of enhancer and promoter) to find each other, In \cite{suppmat} we show this time scale as $n^{1.6..1.67}$ with the genomic distance between the loci, i.e. significantly faster than the Rouse time, enhancing the speed of gene regulation processes. We consider this to be an additional argument in favor of the fractal globule model of genome packing.

Summing up, self-diffusion in a fractal globule state, while much faster than that
in the entangled equilibrium globule, is not described by Rouse model, it is a sub-diffusion with a different exponent $\alpha_F \approx 0.38..0.42$. This result, which we support by scaling theory and computer simulations, is in accordance with earlier numerical \cite{kremer} and experimental \cite{mcl,kepten1,kepten2,kepten3} data. By analogy with the Rouse model, we expect the dynamics in the fractal globule to be a fractional Brownian motion, but full analysis of this matter goes beyond the scope of this letter. Moreover, the compactness of the domains in fractal globule coupled with comparatively fast
subdiffusion leads to the estimate $T \sim n^{1.6..1.67}$ for the first passage time, which is faster than Rouse time $T \sim n^{2}$, not to mention the first passage time in the entangled melt. Ability of different parts of chromatin to find each other fast may be crucial for fast regulation of gene expression. As a bi-product of our simulation we provide evidence that long-living metastable fractal globule state is unique and has characteristics similar to the equilibrium state of non-concatenated polymer rings.

\begin{acknowledgements}
The authors are grateful to V. Avetisov, A. Grosberg, M. Imakaev, S.N. Majumdar, R. Metzler, A. Mironov, S. Nechaev, E. Kepten, A. Semenov, K. Sneppen, and R. Voituriez for many illuminating discussions on the subject
of this work. This work is partially supported by the IRSES project FP7-PEOPLE-2010-IRSES 269139 DCP-PhysBio,
the RFBR grant 14-03-00825, and the Skolkovo Institute of Science and technoogy via SkolTech/MSU Joint
Laboratory Agreement 081-R.
\end{acknowledgements}

\appendix

\section{Supplementary 1. Dissipative particle dynamics}

Dissipative particle dynamics (DPD) is a version of the coarse-grained molecular dynamics adapted to polymers
and mapped onto the classical lattice Flory-Huggins theory \cite{Hoogerbrugge, Schlijper, Espanol, Groot}.
Consider an ensemble of particles (beads) obeying Newton's equations of motion

\be
\frac{d\mathbf{r}_i}{dt}=\mathbf{v}_{i}, {m_i} \frac{d\mathbf{v}_{i}}{dt}=\mathbf{f}_{i}
\ee
\be
\mathbf{f}_i=\sum\limits_{i \neq j} (\mathbf{F}^{b} _{ij}+\mathbf{F}^{c} _{ij}+\mathbf{F}^{d}
_{ij}+\mathbf{F}^{r} _{ij}),
\label{eq:1}
\ee

where $\mathbf{r}_{i}$, ${m}_{i}$, $\mathbf{v}_{i}$ are the coordinate, mass, and velocity of an $i$-th bead,
respectively, $\mathbf{f}_{i}$ is the force acting on it. The summation is performed over all other beads
within the cut-off radius ${r}_{c}$. Below we assume that all quantities entering Eq. (2,3) are dimensionless
and for simplicity set ${r}_{c}$ and ${m}_{i}$ for any $i$ to unity.

First two terms in the sum \eq{eq:1} are conservative forces. Macromolecules are represented in terms of the
bead-and-spring model. $\mathbf{F}^{b} _{ij}$ is a spring force describing chain connectivity of beads:

\be
\mathbf{F}^{b} _{ij}=-K({r}_{ij}-l) \frac{\mathbf{r}_{ij}}{r_{ij}},
\label{eq:2}
\ee

where $K$ is a bond stiffness, $l$ is the equilibrium bond length. If beads $i$ and $j$ are not connected,
then $\mathbf{F}^{b} _{ij}=0$. $\mathbf{F}^{c} _{ij}$ is a soft core repulsion between $i$- and $j$-th beads:

\be
\mathbf{F}^{c} _{ij} = \left\{\begin{array}{ll} a_{ij} (1-r_{ij}) \mathbf{r}_{ij} / r_{ij}, & r_{ij} \leq 1 \medskip \\
0, & r_{ij}>1 \end{array} \right.
\label{eq:3}
\ee

where $a_{ij}$ is a maximum repulsion between beads $i$ and $j$ attained at $r_{i} = r_{j}$. Since
$\mathbf{F}^{c} _{ij}$ has no singularity at zero distance, a much larger time step than in the standard
molecular dynamics could be used.

Other, non-conservative, constituents of $\mathbf{f}_{i}$ are a random force $\mathbf{F}^{r} _{ij}$ and a
dissipative force $\mathbf{F}^{d} _{ij}$ acting as a heat source and medium friction, respectively. They are
taken as prescribed by the Groot-Warren thermostat in \cite{Groot}.

It was shown\cite{Spenley,Lahmar} that the DPD method is consistent with both the scaling theory of polymers
(e.g., it gives correct relationships between the average radius of gyration of a coil and the number of
units in the coil) and the Rouse dynamics.

The use of soft volume and bond potentials leads to the fact that the bonds are formally ``phantom'', i.e.
capable of self-intersecting in three dimensions. The phantom nature of chains does not affect the
equilibrium properties (for example, the chain gyration radius or the phase behavior of the system);
moreover, it greatly speeds up the equilibration of the system. However, the dynamical properties of the
chains, such as the self-diffusion or the features requiring explicit account for the entanglements between
chains, are, of course, dependent on whether the chain is phantom or not, making the situation more subtle.
It is necessary therefore to introduce some additional forces that forbid the self-intersection of the bonds.
These forces are usually quite cumbersome and considerably slow in computation. Nikunen et al. \cite{Nikunen}
described a method for turning chains non-phantom in DPD without introducing any addition forces. It is based
on geometrical considerations: if any two units in the system cannot approach each other closer than
$r_{\min}$, every unit in the system effectively has an excluded radius of $r_{\min}/2$. If it also assumed
that each bond has a maximum length $l_{\max}$, the condition of self-avoiding chains is $\sqrt{2} r_{\min}>
l_{\max}$.

Although particles in DPD are formally point-like, they have an excluded volume due to the presence of the
repulsive potential at any nonzero value of $a_{ij}$. Similarly, the existence of a bond potential causes the
bond to have a maximum possible length.

In our study, we chose $a_{ij} = 150$, $l = 0.5$, and $K = 150$. The other parameters are: DPD number density
$\rho=3$; noise parameter $\sigma = 3$; integration time step $\Delta t=0.03$.

In order to interpret results correctly it is important to make some estimate of the correspondence between the simulation and experimental timescales. One can roughly estimate this correspondence as follows. The chromatin we are aiming to simulate is a DNA-protein complex fiber which is at least 10 nm thick and has a persistence length of the order of 10 nm. Therefore, the size of DPD bead cannot be smaller than 10 nm. Using Einstein's formula
\be
D = \frac{kT}{6\pi \eta R}
\ee 
one gets for the diffusion coefficient of such a bead in water solution at room temperature an estimate of roughly $D \sim 3 \times 10^{-11} m^2/sec$. Therefore, the time at which the bead self-diffuses for a distance equal to its size is of order $ t \sim (bead size)^2 /D \sim 3 \mu sec$. Comparing this result with the results of our simulations (see Figure 3 of the main text) and assuming that for such small displacements the role of chain connectivity is negligible, one gets the estimate of $1 n sec$ per DPD time step. The whole accessible timescale is then of order of $0.1 sec$.

Note that this is an estimate from below, as the simulated media is, generally speaking, more viscous than pure water, and the chain connectivity in fact does play some role in the self-diffusion of DPD beads even on the small time-scales. 

\section{Supplementary 2. Initial states}

In our work we use three different ways to construct initial states of globules which we describe below in
detail. In all cases chain of $2^{18}=262144$ monomers are generated in a cubic box with periodic boundary
conditions, the size of the modeling box is $44 \times 44 \times 44$ reduced DPD units, making the average
number density of monomers equal to $\rho=3$ (this value is known\cite{askgavrilov}) to be especially good
for modeling the dynamic properties of polymer chains). All three initial states are constructed on a cubic
lattice with lattice constant equal to $3^{-1/3} \approx 0.69$ and after the construction are allowed to
anneal for $2^{25} = 3.2 \times 10^7$ DPD time steps. Only after this annealing the self-diffusion
measurements are started.

\subsection{Random fractal globule}

The mechanism of fractal globule formation suggested below is novel and will be discussed and characterized
in full detail in \cite{NazarovinPrep}. Here we provide a brief overview of the idea necessary for the reader
to understand the main text and convince himself that the initial state we are dealing with indeed has all
the properties of a fractal globule.

The idea of this mechanism to design a fractal globule state, which we propose here for the first time, is
based on the following considerations. Imagine a polymer chain being synthesized while being in a poor
solvent, in a way that all the already synthesized part is forming a tight globule. Assume also the synthesis
to be very fast as compared to the internal movements of monomers within a globule. In that way one expects
that at all intermediate stages the already formed part of the globule is in a compact state. Also one expect
that formation of knots and entanglements will be highly suppressed since the new monomers are mostly to the
surface of the existing globule, and cannot go through it as there are no holes left in the structure.
Clearly, the conformation thus formed is very reminiscent of a fractal globule.

\begin{figure}[ht]
\epsfig{file=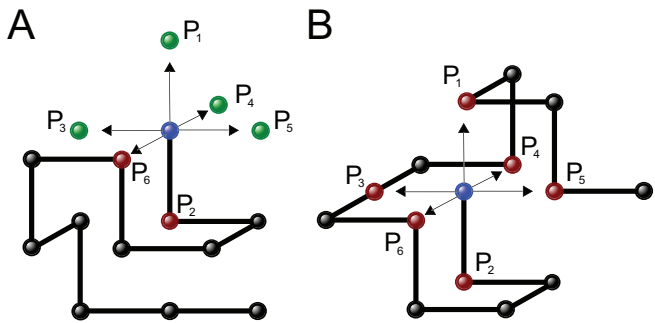, width=8.5 cm} \caption{The conformation-dependent random walk. (A) on the next step
the probabilities of choosing steps 3 and 5 are large compared to steps 1 and 4 ($P_4= P_1; P_3 = (1+2A) P_1
= 20001 P_1, P_5= (1+A) P_1 = 10001 P_1$, probabilities of steps 2 and 6 are proportional to $\varepsilon$
and are essentially zero; (B) trapped configuration: weights of all possible steps equal $\varepsilon$ and
are equiprobable.}
\label{fig:01}
\end{figure}

To exploit this idea we proceed as follows. We construct the polymer conformation as a trajectory of a
lattice random walk in a potential strongly attracting the walker to the places it has already visited. At
each step a walker on a cubic lattice has 6 neighboring cites (see Figure \ref{fig:01}) where he can possibly
move. We postulate the probability to go at each of the possible target cites to depend on whether it was
already visited, and on how many visited cites it has as its neighbors. In particular, we use the following
assumptions:
\be
\begin{array}{rll}
P_{i} & = & \mathcal{N}^{-1} \left\{\begin{array}{l} \varepsilon \medskip \\ \mbox{if the target cite is
visited},
\medskip \\
1 + A\left(\begin{array}{l}\mbox{ \# of visited neighbors}\medskip \\ \mbox{the target cite
have}\end{array}\right) \medskip \\ \mbox{if the target cite is not visited},
\end{array} \right. \medskip \\
\mathcal{N} & = & \sum_{i=1..6} P_i
\end{array}
\ee
Here, $\varepsilon$ should be extremely small so that double visiting of the same cites should be possible
only if the walk gets locked (we use $\varepsilon=10^{-9}$), while $A$ is a constant defining the strength of
attraction to the existing trajectory, and should therefore be large to keep all the intermediate
conformations compact. By trial and error we have found $A=10,000$ to work best.

A trajectory constructed in this way includes a finite fraction (of order of several percent)
self-intersections. However, he resulting states happens to be almost unknotted:

The segment of $10^4$ monomers is reduced to a knot of less than $10^2$ monomers. For comparison, a segment
of an equilibrium globule of $10^4$ monomers is reduced to a knot of $2 \cdot 10^3$ points.

To characterize the resulting states we studied the spatial distance as a function of distance along the
chain, and the return probability of the chain (see \fig{fig:03} and Fig.2 of the main text, see also Fig.2
of the main text for the snapshots of the state). One more important question is whether the characteristics
of the resulting conformations are stable along the chain: indeed, the rules generating the conformation are
asymmetric (the walker is attracted to the sites he visited in the past, but not to the sites he will visit
in the future), so one can expect the start and end parts of the chain to behave differently. To check that,
we calculated the average distance between monomers $\langle R^2 (n) \rangle$ and the return probability
$P(n)$ separately averaged over the first and the last half of the chain. The results are shown in
\fig{fig:startend}. One sees that amazingly there seems to be no bias, and the first and second halves of the
chain behave exactly the same within experimental error.

\begin{figure}
\epsfig{file=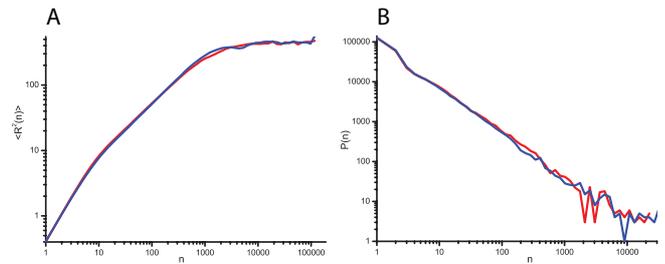, width=8.5cm} \caption{Characterization of the first (red) and second (blue)
halves of the chain generated by the conformation-dependent polymerization algorithm. A) the average spatial
distance between monomers $\langle R^2 (n) \rangle$ as a function of the chemical distance between them $n$,
B) the non-normalized return probability $P(n)$ (number of monomers which happen to be spatially adjacent
while being separated by chemical distance $n$). }
\label{startend}
\end{figure}

The characteristics of thus constructed conformations are, thus, exactly similar to what one expects of a
fractal globule conformations, and moreover, as shown in \fig{fig:03}(A,D) they appear to be very stable
throughout the modelling timescale.

\begin{figure*}[ht]
\epsfig{file=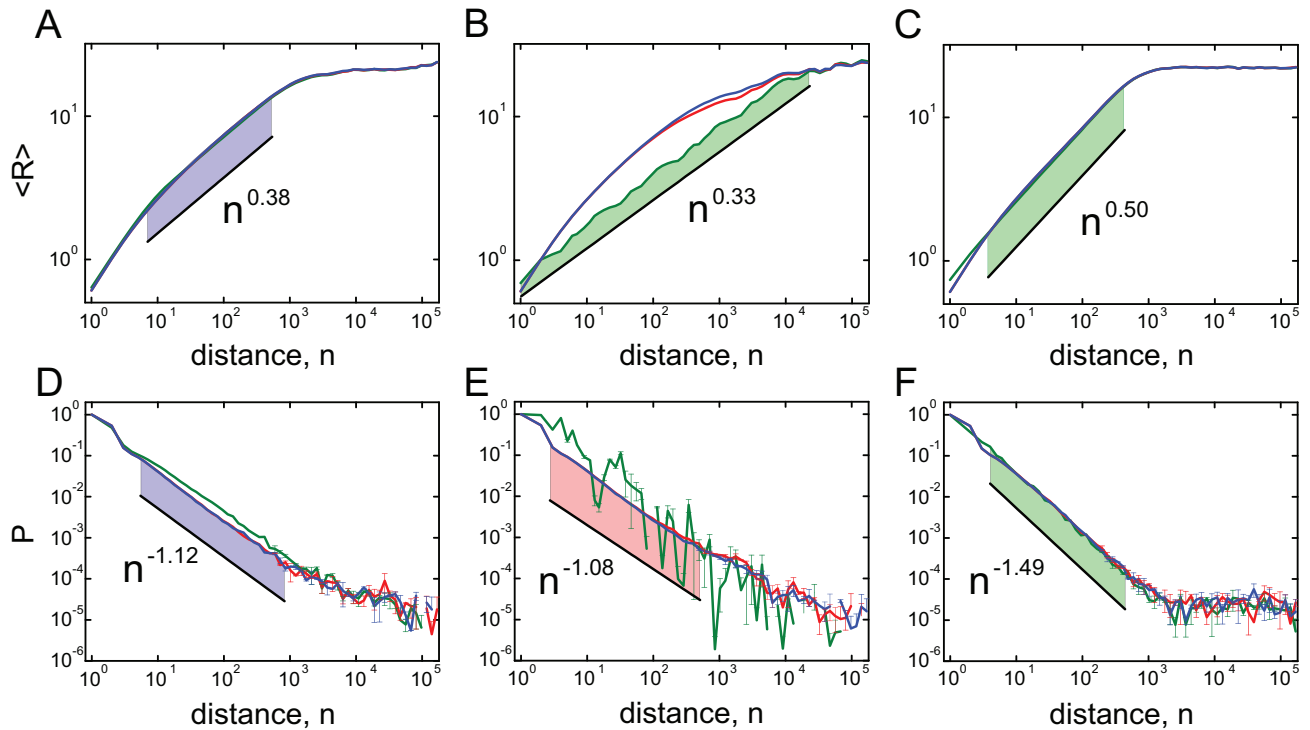, width=17 cm} \caption{The dependence of the average root-mean squared distance
$R(n) = \sqrt{\langle R^2 (n) \rangle}$ between monomers and return probability $P(n)$ on the genomic
distance $n$ between monomers for three initial states under study: (A, D) the random fractal globule, (B, E)
the randomized Moore curve, (C, F) the equilibrium Gaussia globule obtained by the loop-eschange algorithm.
Green curves correspond to initial states, red curves - to the states obtained after annealing for $2^25
\approx 3.2 \times 10^7$ DPD steps, blue curves - to the final states in the end of the modelling, i.e. after
$3 \times 2^25 \approx 9.7 \times 10^7$}.
\label{fig:03}
\end{figure*}

In our opinion, the fractal conformations constructed by conformation-dependent polymerization described
above are not only useful for fast generation of fractal states, but are of fundamental as well as practical
interest. We plan to study them further and in more detail elsewhere \cite{NazarovinPrep}

\subsection{Moore curve}

\begin{figure}
\epsfig{file=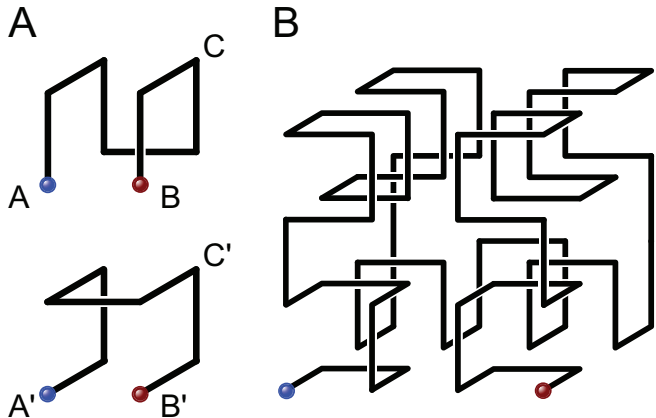, width=8.5 cm} \caption{The space-filling curve design: (A) first iteration of the
construction. There are two mirror-symmetrical ways to circumvent the $2 \times 2 \times 2$ designated ACB
and A'C'B', respectively; (B) second iteration of the construction: ABC curve is swollen by a factor of two,
each node replaced by a $2 \times 2 \times 2$ subcube and all subcubes are circumvented according to the
A'B'C' curve.}
\label{fig:02}
\end{figure}

Moore curve is an example of a class of recursively defined space-filling curves whose various definitions go
back to the end of 19th century\cite{Peano, Hilbert}. On the first iteration step consider a curve passing
through the cube vertexes as shown in \fig{fig:02}(A). The next iteration consist of i) swelling up the 1-st
iteration curve by a factor of 2, ii) replacing each ``swollen'' vertex of the curve with a $2 \times 2
\times 2$ cube, iii) filling up this newly formed cubes by replicas of the original 1-st iteration curve
rotated in a way to preserve the connectivity of the curve as a whole (see \fig{fig:02} (B)). Then this
procedure can be repeated again and again - swelling up the existing curves and replacing each node with a $2
\times 2 \times 2$ sub-cube whose nodes are circumvented by a curve identical to the 1st iteration curve
rotated in the way to preserve overall connectivity of the picture.


Different ways of constructing space-filling curves (there are Peano version, Hilbert version, etc.) differ
by the exact details of the iteration procedure. Note that in the Moore procedure described above there are
two mirror-symmetrical ways to choose the initial curve of first iteration, as shown in \fig{fig:02}(A).
Consequently, there exist some ambiguity in choosing the particular configuration: indeed, on each iteration
step, when replacing a node with a fragment of curve visiting vertices of a $2 \times 2 \times 2$ subcube it
is possible to choose (independently each time) one of the two mirror-symmetrical ways to circumvent the
nodes.

In simulations we deal with two realizations of a space-filling curve Moore curves formed after 6 iterations
and consisting of $(2 \times 2 \times 2)^{6}=262144$ nodes. One conformation is a ``stereo-regular'' Moore
curve with some particular choice of mirror-symmetric variants. Another allows the choice of ``stereo
isomers'' shown in \fig{fig:02}A at random (this procedure is very similar to one described in
\cite{Fractalglobule_modeling}.  This leads to a conformation with same statistical characteristics but
different intra-chain contacts. We have not noticed any measurable difference in the long-range statistics
and evolution of these states, so under label `Moore' in the main text we provide the results averaged over
these two realizations.

Remarkably, as these initial states anneal, their characteristics seem to relax towards those of a fractal
globule state described above, see \fig{fig:03}(B,E).

\subsection{Equilibrium globule}

\begin{figure}
\epsfig{file=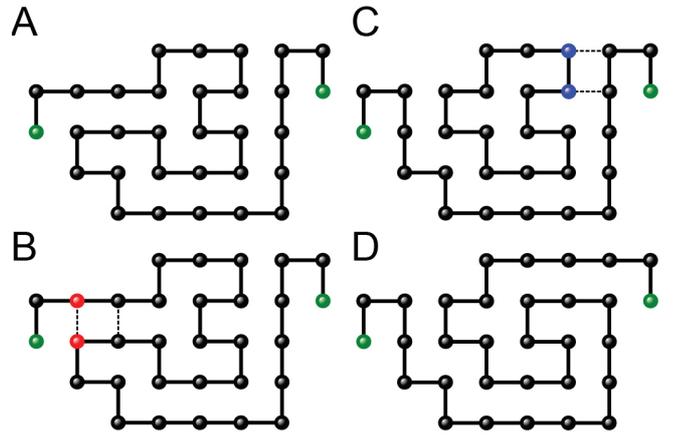, width=8.5 cm} \caption{The scheme of a loop-transmission algorithm. A) The initial
conformation of the chain; B) first bond switching, formation of a separate loop; C) intermediate state with
a separate loop and second bond switching; D) the resulting polymer conformation.}
\label{fig:04}
\end{figure}

Since the full relaxation time of a globule goes far beyond the time-scales available at our simulations, we
have to use some special trick to obtain a reference equilibrium globule conformation with Gaussian
statistics of the short chain fragments. To do that we proceed as follows. We start with a Moore curve
described in the previous subsection and then let it undergo a loop transfer algorithm as follows. At each
step of the algorithm we i) take an arbitrary pair of units which are adjacent on the lattice but not along
the chain (shown in red in \fig{fig:04}(B)) and switch the links between these units in a way showin in
\fig{fig:04}(B), as a result the chain fragment between these two units is ``cut out'' and form a separated
loop. Then, ii) we go along this newly formed loop and choose an arbitrary link of it (between monomers shown
in blue in \fig{fig:04}(C) which is adjacent and parallel to a link of the original chain and perform another
links switch as shown in \fig{fig:04}(C) to insert the loop into a different place of the original chain.
Such a loop transfer preserves the total chain length and the uniform density over the entire volume, but
allows a significant change of the conformation topology. After repeating this procedure many times (of the
order of the total number of chain links, by trial and error we found that $5 \times 10^4$ successful
transfer steps is sufficient for our chain-length) we obtain a lattice conformation which is connected and
uniform in space but otherwise completely random. Indeed, one can see that the dependence of an average
spatial distance between two monomers $\langle R(n) \rangle$ on a distance between them along the chain $n$
is transformed into the dependence which is typical for the equilibrium globule state, i.e.
\be
\begin{array}{rll}
\langle R(n) \rangle & \sim n^{1/2} & \mbox{  for  } n < N^{2/3}; \medskip \\
\langle R(n)\rangle & \sim  \mbox{const} &\mbox{  for  } n > N^{2/3},
\end{array}
\ee
see \fig{fig:03}(C), the contact probability $P(n)$ also changes to one typical for Gaussian chains:
\be
\begin{split}
P(n) \sim n^{-3/2}\mbox{  for  } n < N^{2/3}; \\
P(n) \sim \mbox{const}  \mbox{  for  } n > N^{2/3},
\end{split}
\ee
see \fig{fig:03}(F).

Throughout the modelling the statistical properties of thus obtained equilibrium knotted state remain
constant (see \fig{fig:03}(C,F)).

We find this loop transfer algorithm of producing an equilibrium knotted state with a uniform density in a
cubic volume to be very simple, fast and useful.

\section{Supplementary 3. The role of hydrodynamic interactions}

In this section we discuss how the results for the self-diffusion in a fractal globule obtained in the main text, namely the Equation (7), can be modified to allow for the the hydrodynamic interactions.
The central idea of the Rouse-type approach used in the main text is to assume that the monomers' friction versus surrounding medium is independent, and, therefore, the effective diffusion coefficient of a blob consisting of $N$ monomers is inversely proportional to $N$.
One may argue that such an independent friction is not realistic, especially given that a fractal globule domain is a compact object so that its inner monomers see nothing but the monomers of the same domain around them, so there cannot be any friction between them and the medium of the outer space. Following this reasoning it seems natural to assume that only surface monomers of a blob undergo the friction. Since the surface of a blob is highly fluctuating, it seems natural to assume that the friction of these surface monomers is independent, which leads to a $\tau$-dependent diffusion
coefficient to be inversely proportional to the number of monomers on the surface of the blob:
\be
D(\tau) \sim \frac{D_0}{A(\tau)/A_0},
\label{Rouse}
\ee
(compare with Eq. 4 of the main text), where $D_0$ and $A_0$ are the microscopic parameters (roughly the diffusion coefficient and a surface area of a single monomer unit), and $A(\tau) \sim A_0 [\delta s(\tau)]^{\beta}$ is the surface area of the blob. The exact value of exponent $\beta$ has been recently discussed in the literature. Theoretically, simple mean-field
calculation gives\cite{Lieberman} $\beta =1$, while a more sophisticated analysis leads to a conjecture
\cite{GrosbSoft} $\beta \approx 0.91$. The numerical results seem to suggest it to be slightly below 1,
$\beta \approx 0.93 \pm 0.02$ \cite{HalvJCP,nechaev} (although, see \cite{WittmerJCP} where this result is
challenged). Accordingly, in what follows we assume $\beta$ to belong to the $0.91<\beta<1.$ range.

This generalization yields, instead of equation (7)
\be
\begin{array}{rll}
<(x(s,t+\tau)-x(s,t))^2> & \sim & (\tau)^{\alpha} \sim \frac{D_0}{\delta s^{\beta}} \tau \sim \tau^{1-\beta \alpha d_f/2}; \medskip \\ \alpha_{F} &= & \frac{2}{2+\beta d_f}\simeq 0.42, 
\end{array}
\label{Rouse_alpha} 
\ee
for a model allowing for the hydrodynamic interactions. Note, however, that the shift from the prediction of the Rouse-like theory, $\alpha_F = 0.4$ is minor, and is in the direction opposite from that of computer simulation data. We expect, therefore, that fluctuation effects in this system play more significant role than the hydrodynamic interactions.

\section{Supplementary 4. Distribution of monomer displacements}

\begin{figure*}[ht]
\epsfig{file=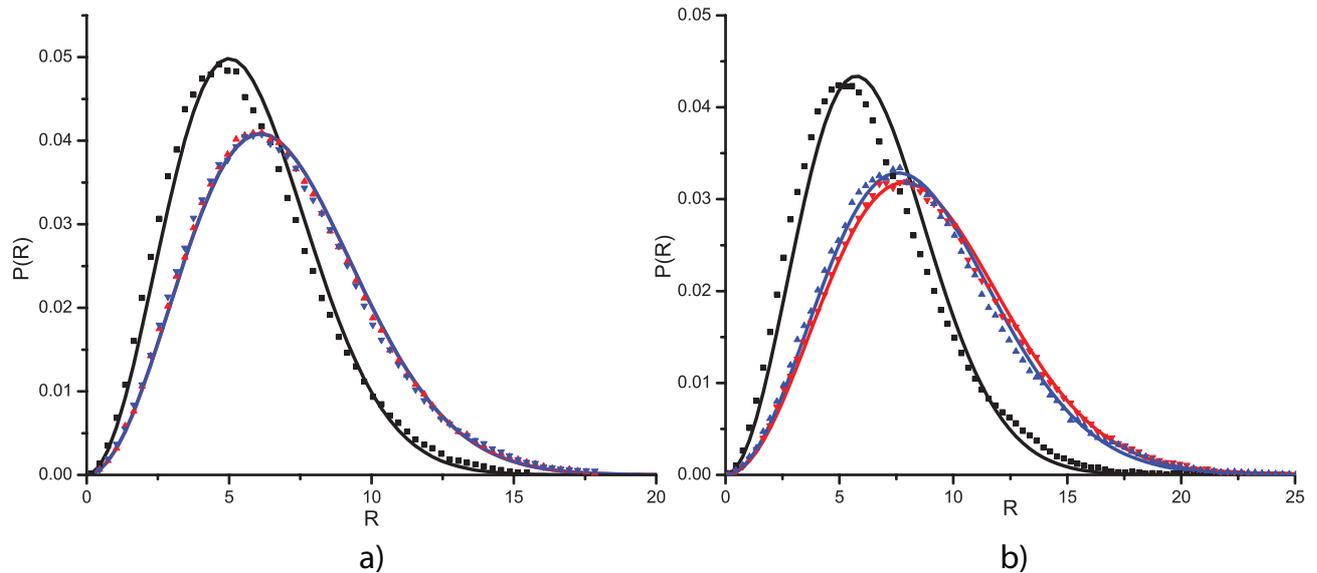, width=17 cm}
\caption{Distribution of the monomer displacements from initial states up to (a) $1.6 \times10^7$, (b) $6.5 \times 10^7$ timesteps. The initial states are equilibrium globule (black squares), annealed Moore (red triangles) and annealed random fractal (blue triangles). The best fits by Maxwell distribution are shown with full lines of corresponding colors.}
\label{fig:dist}
\end{figure*}

To get a better insight into the properties of the chain movement, we have studied the distribution of the monomers displacements from their original positions. We show corresponding results for the distribution of the absolute value of monomer displacements in \fig{fig:dist}. Note that if the displacements are normally distributed, their absolute value should obey the Maxwell distribution:
\be
P(R) = \frac{3 \sqrt{6}}{\sqrt{\pi}(<R^2>)^{3/2}} R^2 \exp \left( -\frac{3R^2}{2<R^2>}\right)
\label{Maxwell} 
\ee 
The corresponding dependences are shown with full lines in \fig{fig:dist}. One sees that while the displacement distributions for equilibrium globule show visible deviations from the Gaussian behavior, those for the fractal globule states (both with Moore and random fractal initial states) seem to stick to the normal distribution. In our opinion, this behavior may indicate that the underlying process can be well described by a fractional Brownian motion with a proper Hurst index $H = \alpha_F/2$. However, this hypothesis needs further confirmation. In particular, one needs analyze individual trajectories of the particles to check that self-diffusion in a fractal globule is ergodic \cite{metzler_review}. Such analysis, however, is going beyond the scope of this article and will be provided elsewhere.

\section{Supplementary 5. First passage time estimation}

The theory presented in this latter can be used to estimate the first passage time for finding a fixed target
within a globule. Since $2/\alpha_F>d$ ($d$ here is the dimensionality of the underlying space, not to be
confused with $d_f$), we expect the sub-diffusion of monomer units to be recursive \cite{Hughes,voituriez},
and the volume of the space visited by one monomer to grow as
\be
V(t) \sim \left( \langle X^2(t) \rangle \right)^{d/2} \sim t^{d \alpha_F/2}.
\label{volume}
\ee
Therefore the typical time needed to explore the whole volume of the globule formed by the chain of $N$
monomer units (which is needed to find a microscopic randomly-placed target) scales as
\be
\begin{array}{rll}
\disp T_{target} & \sim V^{2/(d \alpha_F)} \sim N^{(2+\beta d_f)/d_f} \medskip \\
(2+\beta d_f)/d_f & \approx 1.6
\end{array}
\label{FPT}
\ee
Similarly, the typical cyclization time for a subchain of length $n$ (i.e., the typical time needed for the
two units separated by a subchain of length $n$ to meet each other in space for the first time, see
\cite{wilfix,likhtman,benichou}) equals the time needed to explore the volume $v \sim n^{d/d_f}$ and is given
by
\be
T_{cycl} \sim n^{(2+\beta d_f)/d_f} \approx n^{1.6}
\label{FPT2}
\ee
This time is much smaller than the typical Rouse time $T_R=n^2$, which regulates cyclization of a Gaussian
polymer coil in a viscous environment, not to mention the typical time needed for the units of an entangled
globule to find each other (in the entanglement-controlled case $N_e<n<N^{2/3}$ it can be estimated as
$T_{\mbox{ent}} \sim n^{ 2/(d_f \alpha_{\mbox{ent}}} \sim n^4$, where $d_f=2$ due to the Flory theorem
\cite{GrKhokh}). Therefore, we see that fractal globule state is in this sense advantageous: it takes
monomers less time to find each other in this state, thus enhancing the speed of gene regulation processes.
We consider this to be an additional argument in favor of the fractal globule model of genome packing.

\end{document}